\documentclass[
]{ceurart}

\usepackage[frozencache=true,cachedir=minted-cache]{minted}
\usepackage{CJKutf8}

\newcommand{\chinese}[1]{\begin{CJK}{UTF8}{bsmi}#1\end{CJK}}

\setminted{breaklines=true}

\begin{document}

\copyrightyear{2021}
\copyrightclause{Copyright for this paper by its authors.
  Use permitted under Creative Commons License Attribution 4.0
  International (CC BY 4.0).}

\conference{CLEF 2021 -- Conference and Labs of the Evaluation Forum, 
	September 21--24, 2021, Bucharest, Romania}

\title{Recognizing bird species in diverse soundscapes under weak supervision}

\author[1,4]{Christof Henkel}[%
email=chenkel@nvidia.com,
]

\address[1]{NVIDIA, Munich, Germany}

\author[2,4]{Pascal Pfeiffer}[%
email=pascal.pfeiffer1@rwth-aachen.de,
]
\address[2]{RWTH Aachen University, Aachen, Germany}

\author[3,4]{Philipp Singer}[%
email=philipp.singer@h2o.ai,
]
\address[3]{H2O.ai, Mountain View, USA}

\address[4]{All authors contributed equally.}
\begin{abstract}
We present a robust classification approach for avian vocalization in complex and diverse soundscapes, achieving second place in the BirdCLEF2021 challenge. 
We illustrate how to make full use of pre-trained convolutional neural networks, by using an efficient modeling and training routine supplemented by novel augmentation methods. 
Thereby, we improve the generalization of weakly labeled crowd-sourced data to productive data collected by autonomous recording units. 
As such, we illustrate how to progress towards an accurate automated assessment of avian population which would enable global biodiversity monitoring at scale, impossible by manual annotation. 
\end{abstract}


\maketitle

\vspace{-1em}
\section{Introduction}
\label{sec:introduction}

Avian species are important members of our ecosystem providing regulating services like pollinating plants, dispersing seeds and controlling pests. To understand the functioning of a local ecosystems, it is important to detect and identify individual bird species. Since manual monitoring of avian species is impossible at a large scale it is crucial to develop autonomous systems for monitoring and classification.

The yearly Bird Recognition challenge (\emph{BirdCLEF}) puts attention on the reliable detection of avian vocalizations in various soundscapes. The \emph{BirdCLEF2021} edition \cite{birdclef2021} as part of \emph{LifeCLEF2021} \cite{lifeclef2021, plantclef2021, snakeclef2021, geolifeclef2021} especially focuses on generalization of audio data from the crowd-sourced bird sound database \emph{xeno-canto} \cite{xenocanto} towards analyzing complex and long soundscape recordings generated in different contexts. 

The goal of this year's competition is to predict the presence of individual bird species in five second intervals for recordings across four distinct locations in North and South America. Contrarily, the provided training data only contains short recordings of individual bird calls extracted from user uploads on the xeno-canto platform. Consequently, competitors had to address a weak label problem to bridge the gap between training and testing data discrepancy.

In this work, we discuss an efficient model architecture and training routine as part of the second place solution of the BirdCLEF2021 competition. 
Our solution builds upon a robust bootstrapped validation setup, and consists of an ensemble of convolutional neural networks (CNN) trained on extracted spectrograms of short audio clips. Our developed modeling architecture successfully bridges the gap between the weakly labeled crowd-sourced training data and the more accurately hand-labeled test data with hard labels at five second intervals. To further account for label noise in training data, we employ several augmentation and training methods such as novel mixup variations, the addition of background noise, or label smoothing. 
Final predictions are supplemented by a binary bird presence classifier and several post processing steps.
Implementation details can be found on github\footnote{\label{githubrepo}\url{https://github.com/ChristofHenkel/kaggle-birdclef2021-2nd-place}}.

\newpage
\section{Material}
\label{sec:material}


\subsection{Competition Data}

The final test set of this competition includes approximately 80 soundscape recordings across the four unique locations
Columbia (COL), Costa Rica (COR), Sierra Nevada (SNE) in California, USA and Sapsucker Woods
area (SSW) in New York, USA.
Each recording lasts ten minutes and predictions had to be provided for each five second interval of each audio file addressing the presence or absence of individual bird species. Additional meta data including the time of recording as well as the approximate coordinates of the location is provided for the test data.

Contrarily, the training data only contains weak labels for around $63,000$ short recordings of individual bird calls shared by users on the xeno-canto platform. Overall, $397$ bird species are captured in this training set, out of which only an unknown set is also present in the test set. For each recording, a primary label is specified, as well as an optional list of secondary labels. Additional meta data including the date of recording, geo-location, the user uploading the data, as well as a quality rating from $0-5$, where $5$ indicates best quality, is present. 

To allow for better local validation of models, the competition data also includes a separate validation soundscape dataset which is similarly structured as the final test dataset. However, it includes only $20$ soundscape files across two out of the four test data locations and does not necessarily capture the same bird species as the final test set. Yet, it allows to roughly bridge the gap between the difference of weak labels used for training, and hard labels used for testing, and allows for better validation as discussed in Section~\ref{sec:methods}.

\subsection{External Data}

To improve the generalization of any model trained on the crowd-sourced database xeno-canto and apply it to new unseen data we augmented the training data with background noise not containing avian calls. 
In particular, we used two additional external datasets \textbf{freefield1010} \cite{freefield1010} and \textbf{BirdVox-DCASE-20k} \cite{birdvox} which had binary labels for the presence/absence of bird sounds and were advertised in the DCASE2018 Challenge \cite{Stowell_2018badchj}. Moreover, we extracted consecutive 30 second clips containing no avian sound from the validation set of the previous BirdCLEF2020 competition \cite{birdclef2020} which was hosted on aicrowd \footnote{\url{https://www.aicrowd.com/challenges/lifeclef-2020-bird-monophone}}.

\section{Methods}
\label{sec:methods}



\subsection{Validation}
In Kaggle code competitions, the test dataset is hidden in such a way, that only very limited information can be gathered about it other than the public leaderboard score, which resembles the score for 35\% of the total test data described in Section~\ref{sec:material}, also called \emph{public test data}. The final score is calculated on the remaining 65\% of the data, also called \emph{private test data}. In this particular machine learning competition, the training data is very different to the test data, as the audio files only contain weak labels without time information and differ in length. Nevertheless, for two of the four test sites, a small validation set resembling the test set is provided by the hosts of the competition to give competitors a general understanding of the test data's nature. The metric of choice according to which competitors are evaluated in this competition is the row-wise micro averaged F1 score.

Consequently, our models are only trained on the short training clips, and locally evaluated on the provided validation soundscapes.
This validation scheme was further improved by omitting three full soundscapes (out of 20) that did not contain any bird calls at all. In the following, this validation scheme is referred to as CV-3.
To receive an even more robust feedback from validation, a bootstrap sampling method was introduced. 
Our final validation setup included the following steps:

\begin{itemize}
\item Remove three songs without calls.
\item For k times (e.g., $10$) sample $80\%$ of the remaining songs---this should emulate the full test dataset (public+private).
\item Apply any kind of threshold selection technique or post processing on this data resembling the application in inference for the full public and private test data.
\item For j times (e.g., $50$) sample $65\%$ of the remaining songs---this should emulate the private test dataset.
\item Calculate the score on each of these j samples.
\item Report average, median, min, max, std scores across all k times j (e.g., $500$) subsets
\end{itemize}


\subsection{Modeling}

Our final solution is a an ensemble of CNNs which were trained on extracted spectrograms of the audio clips. Their predictions were further post-processed using a binary CNN for classifying the presence/absence of avian sound and available meta data with respect to date and region of the test soundscapes. In the following, we discuss design and training routine of the CNNs. 
All our models were trained on GPUs utilizing the Pytorch framework and backbones that were 
pre-trained on ImageNet and that are available at timm \cite{rw2019timm}. In order to perform computational efficient training, we also performed the spectrogram transformation on GPU using torchaudio\footnote{\url{https://pytorch.org/audio/stable/index.html}} and mixed precision.

\subsubsection{Bird Classifier}

The core part of our solution addresses the main task-specific challenges: Firstly, training data is weakly labeled with bird sound classes being annotated on audio clips of varying length, while prediction needs to be performed on five second time windows of several minute long recordings. Secondly, due to the complex soundscapes and broad range of data contributors, the training labels contain a significant amount of low quality annotations. Lastly, the test data is of significantly different context and of different quality.

\begin{figure}
  \centering
  \includegraphics[width=\linewidth]{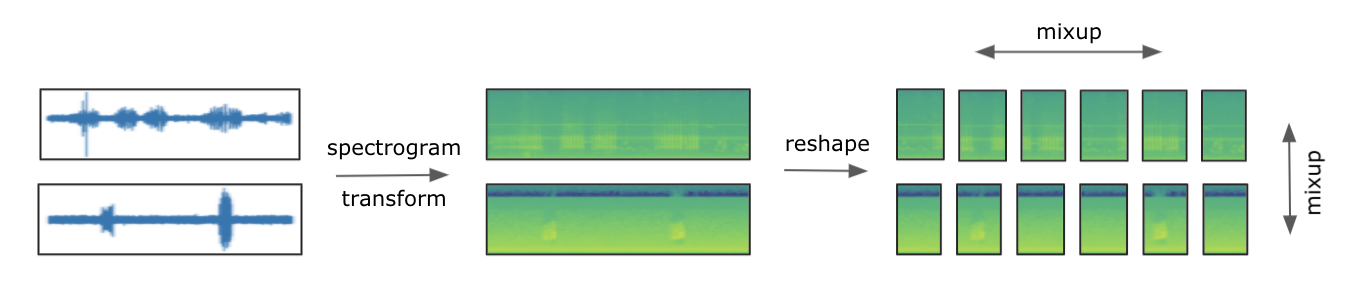}
  \caption{Visualization of preprocessing steps for the bird classification model --- Input is a 30 second wav crop. After mel-spectrogram transformation, the input data is split into six equal five second parts, before mixup augmentation is applied within and between recordings.}
  \label{preprocessing_figure}
\end{figure}

\paragraph{Preprocessing}
The training data contains short sound clips of different length with weak primary and secondary labels. To preprocess the training data for modeling, we randomly cropped a 30 second time window from a recording and applied a mel-spectrogram transformation. Hereby, 30 seconds proved to be a good compromise between label accuracy and generalization provided by the augmentation. To account for the five second time window required for inference of the test data, we reshaped the crops into six parts. Before feeding the spectrograms into the CNN-backbone, we applied a two dimensional version of mixup \cite{mixup}. In particular, we not only mixed between different recordings (up to two times), but also within a recording by mixing the six parts. Our preprocessing was performed on GPU using functionalities of torchaudio for mel-spectrogram transformation and a custom implementation for mixup. We illustrate our preprocessing pipeline in Figure \ref{preprocessing_figure}. 

\begin{figure}
  \centering
  \includegraphics[width=\linewidth]{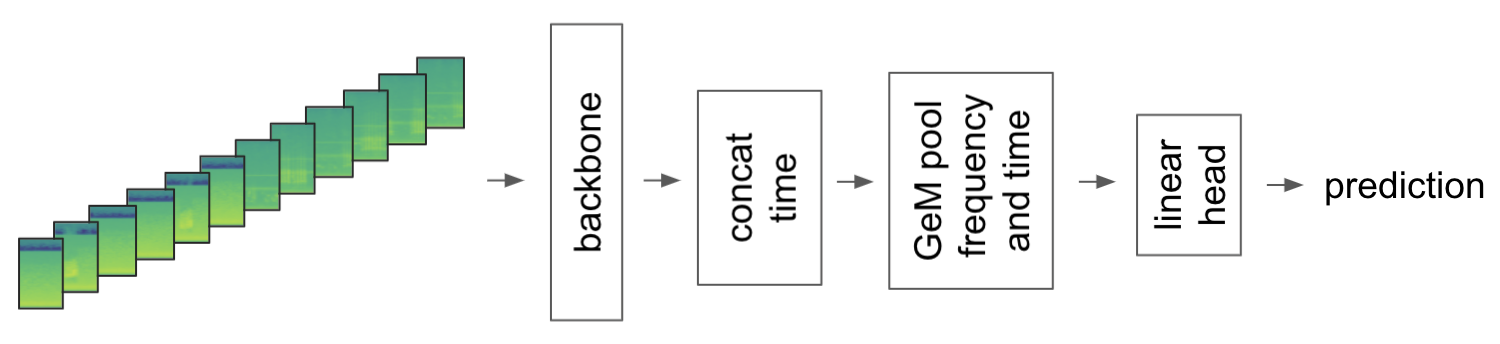}
  \caption{Illustration of bird classification model --- five second mel-spectrograms are fed into a pretrained CNN-backbone and re-arranged to 30 second time segments before GeM pooling is applied. A single linear layer outputs the bird classes.}
  \label{bird_arch_figure}
\end{figure}

\paragraph{Architecture}
Given a batch of mel-spectrograms, the model extracts features with a CNN-backbone pretrained on ImageNet, where in our final solution we used backbones from the resnet \cite{resnet} efficientnet v2 \cite{efficientnetv2} and nfnet \cite{nfnet} family, which are all available in the timm repository \cite{rw2019timm}. We re-arrange the tensors to the 30 second representation by concatenating the respective time segments and use generalized mean pooling (GeM) \cite{gem} of time and frequency dimension before forwarding through a simple one layer head which results in a prediction of 397 bird classes. For inference we directly fed five second snippets to the CNN-backbone and applied the head without reshaping.
See Figure \ref{bird_arch_figure} for an illustration.

\paragraph{Training routine}

We did not differentiate between primary and secondary labels and set the multilabel target as the union of both.
We trained our models with a binary cross-entropy loss using Adam optimizer and a 11 to 20 epochs lasting cosine annealing schedule with a batch size varying from 16 to 32 per model. While the model is not very sensitive to hyperparameters of optimizer and learning rate schedule, several adjustments lead to significant improvement in classification quality.

Firstly, we used the quality rating, which is given as meta information for the training data for weighting a recording's contribution to the loss. The assumption is that recordings with a lower rating have worse quality with respect to audio and label and should contribute less to model training. In detail, we weight each sample by rating/max(ratings).

Secondly, we used one-sided label smoothing, i.e., adding $0.01-0.025$ across all negative labels while positive class is unchanged. This accounts for noisy annotations and absence of birds in ``unlucky`` 30 second crops.
Similar to solutions of past BirdCLEF challenges, we mixed the training data with additional background noise containing no avian calls. Specifically, we used recordings from the freefield dataset labeled for bird absence, three recordings of this years validation set containing no bird calls and 30 second crops of past years validation data containing no bird calls according to their annotations; for details, please refer to Section~\ref{sec:material}.

\subsubsection{Binary classifier}

A binary classifier was trained to distinguish any bird call from other environment sounds and predictions were used in the ensembling/post-processing step together with the bird classifier models. For this task, we only used external data, the binary labeled \textbf{freefield1010} \cite{freefield1010} and \textbf{BirdVox-DCASE-20k} \cite{birdvox} datasets, both containing ten second audio recordings with or without avian sounds (see Section~\ref{sec:material}).

\begin{figure}
  \centering
  \includegraphics[width=\linewidth]{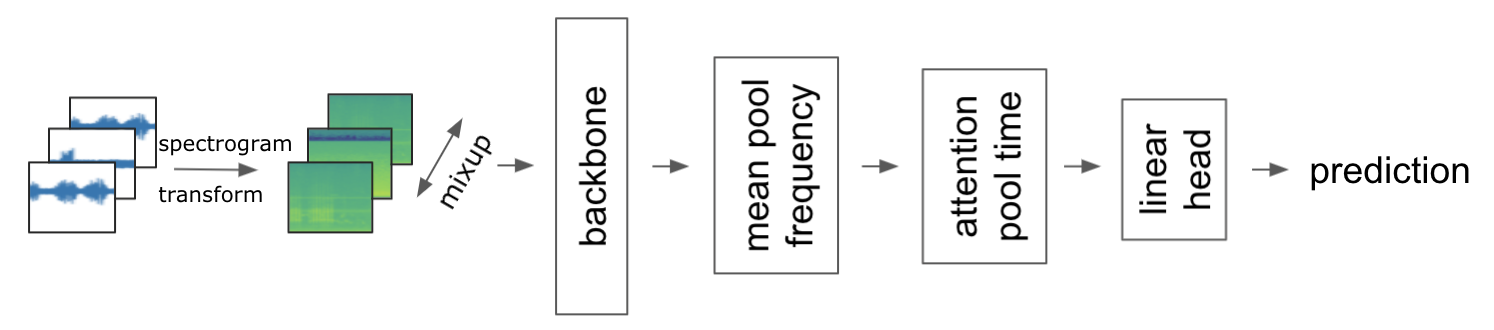}
  \caption{Illustration of binary classification model --- Ten second recordings are transformed to spectrograms before augmented by mixup. Using a pre-trained CNN-backbone features are extracted which are pooled by mean and attention before a single linear layer outputs a binary bird presence / absence prediction}
  \label{binary_arch_figure}
\end{figure}

Figure \ref{binary_arch_figure} gives an overview of the binary classifier. After transforming the recordings to mel-spectrograms, mixup augmentation is applied before feeding into a pretrained CNN-backbone. The output feature map is mean pooled on the frequency axis before applying attention pooling on time segments. A single linear layer is applied to output a binary bird presence /  absence prediction.

\subsection{Postprocessing}
\label{subsec:pp}

As F1 metric is evaluated on hard predictions without taking probabilities into account, appropriate thresholds for detections in a five second segment in soundscapes had to be found. Optimizing the thresholds on CV-3 and applying it to the test set already indicated good correlation. Nevertheless, submitting to the public leaderboard with lower thresholds (higher number of predictions) yielded even better F1 scores. Additionally, one major issue we observed was when blending new models, the threshold needs to be re-optimized each time. 
Hard thresholds are not necessarily consistent across models, as each one can exhibit certain shifts in probabilities and their averages. Consequently, it is challenging to properly judge if new models work well in an ensemble based on the merit of the models, or only based on arbitrary probability / threshold shifts that emerged from it. For a more robust approach regarding shifts in model probabilities, we decided to apply a percentile based thresholding approach. In detail, we flattened all predictions and calculated the threshold for a certain percentile which we optimized independently for maximum F1 score on CV-3 and on public leaderboard.


The more birds a set contains, the lower the percentile should be if predictions are decently ranked. With this approach, we kept the percentile fixed, thus always predicting the same amount of bird calls, and just exchanged models, blends and other post processing steps. If the quality in our ranking of predictions improved, also the score improved.

On our bootstrapped CV-3, we found a percentile of $0.9987$ to be optimal. Based on public leaderboard feedback, we decided to decrease the percentile a bit further to $0.9981$ for our test predictions expecting a few more birds to be predicted. In retrospective, a slightly higher percentile would have further improved our solution.
By and large, our percentile based thresholding approach appeared to be robust across CV and LB and generalized well to the unseen private part of the test set.

Additionally, we employed several smaller post processing steps to improve predictions including: (1) increasing the probability of birds in songs based on their average prediction probability, (2) smoothing neighboring predictions, and (3) we also removed some unlikely predictions based on distance in space and time given the provided metadata. Additionally, we utilized our binary models predicting whether a bird song is present in a given window or not and adjusted our predictions by $p_{bird}=p_{bird}*(1+p_{binary}*0.8)$. Overall, we fitted 15 binary models and averaged them.

\subsection{Ensembling}

The ensembling \cite{hansen1990neural} of our models was straightforward since all outputs have the same shapes; we took the arithmetic mean of the predictions.
Post-processing steps (1) and (2) as described in Section~\ref{subsec:pp} were applied individually on each model, all other steps were applied after blending. Our final solution contains nine models differing only on hyperparameters and backbones. Each type of model was fitted with six different seeds accounting for variability in predictions. Our final inference Kaggle kernel\footnote{\url{https://www.kaggle.com/ilu000/2nd-place-birdclef2021-inference/}} ran only one hour out of three hours allowed giving us further potential room for improvements. In the following Section~\ref{sec:results}, we describe according results and go into more detail about the individual models.

\begin{figure}
  \centering
  \includegraphics[width=0.7\linewidth]{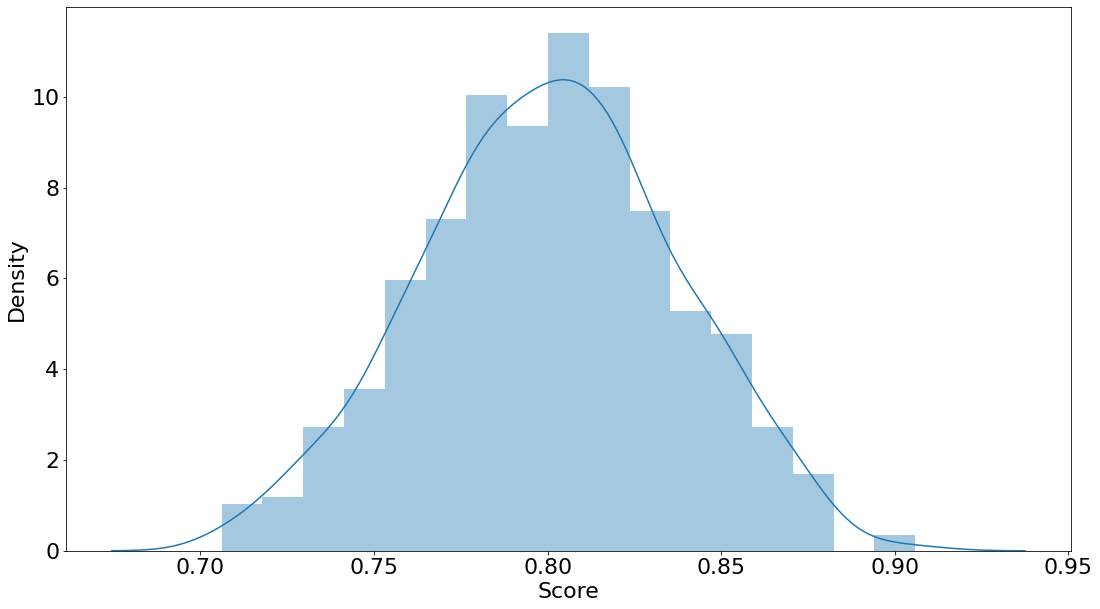}
  \caption{Visualization of bootstrapped validation results --- This figure depicts the distribution of scores over 500 iterations based on our bootstrap validation routine.}
  \label{fig:cv_results}
\end{figure}

\begin{table}[t!]
\caption{Model results --- This table highlights individual model and final ensemble results sorted by CV score. All models exhibit same postprocessing steps, but differ in utilized backbones and hyperparameters. The main hyperparameter differences stem from melspec settings, background augmentation, and label smoothing (LS). For melspec settings, S1 refers to window\_size=1024, hop\_size=320, fmin=50, fmax=14000, mel\_bins=256, power=2, and top\_db=80, while S2 refers to window\_size=2048, hop\_size=512, fmin=16, fmax=16386, mel\_bins=64, power=2, and top\_db=None.}
\begin{tabular}{| l | l | c | c | c | c | c | c |}
\hline
Model & Backbone & Melspec & \makecell{Back \\ ground} & LS & CV & Public LB & Private LB \\ 
\hline
ps-6-v2 & resnet34 & S1 & no & 0.000 & 0.7387 & 0.7649 & 0.6592 \\
ps-12-v8 & resnet34 & S1 & yes & 0.000 & 0.7698 & 0.7665 & 0.6832 \\
ps-12-v13 & tf\_efficientnetv2\_m\_in21k & S2 & yes & 0.000 & 0.7716 & 0.7610 & 0.6719 \\
ps-12-v21 & eca\_nfnet\_l0 & S2 & yes & 0.000 & 0.7803 & 0.7892 & \textbf{0.6846} \\
ps-12-v11 & tf\_efficientnetv2\_s\_in21k & S1 & yes & 0.000 & 0.7807 & 0.7848 & 0.6751 \\
ps-12-v30 & tf\_efficientnetv2\_s\_in21k & S2 & yes & 0.025 & 0.7861 &  0.7841 & 0.6667 \\
ch-12-v25a & tf\_efficientnetv2\_s\_in21k & S2 & yes & 0.010 & 0.7901 &  \textbf{0.7917} & 0.6785 \\
ps-12-v32 & tf\_efficientnetv2\_s\_in21k & S2 & yes & 0.010 & 0.7855 &  0.7838 & 0.6656 \\
ch-12-v25g & tf\_efficientnetv2\_s\_in21k & S2 & yes & 0.010 & \textbf{0.7975} &  0.7869 & 0.6845 \\
\hline
Ensemble  & & & & & \textbf{0.8000} &  \textbf{0.7998} & \textbf{0.6893} \\
\hline
\end{tabular}
\label{tab:results}
\end{table}

\begin{table}[b!]
\caption{Leaderboard results --- This table shows the ranking and scores of the top five teams on the private leaderboard including their respective public leaderboard scores. The team of this article is highlighted in bold.}
\begin{tabular}{| l | l | l |}
\hline
Team & Public LB & Private LB \\ 
\hline
Dr.\chinese{北村の愉快な仲間たち} & 0.7736 (\#17) & 0.6932 (\#1) \\
\textbf{new baseline} & 0.7998 (\#3) & 0.6893 (\#2) \\
Shiro & 0.7919 (\#4) & 0.6891 (\#3) \\
Third time's the charm. & 0.7897 (\#6) & 0.6864 (\#4) \\
Kramarenko Vladislav & 0.7897 (\#7) & 0.6820 (\#5) \\
\hline
\end{tabular}
\label{tab:lbresults}
\vspace{-1em}
\end{table}

\section{Results and discussion}
\label{sec:results}

As described in Section~\ref{sec:methods}, our final solution is based on an ensemble of multiple models with different hyperparameter settings and seeds. In Figure~\ref{fig:cv_results}, we visualize our bootstrapped CV-3 validation results; the x-axis depicts the score, and the y-axis respective density across all 500 bootstrap iterations. The distribution exhibits a mean of $0.800$, a median of $0.801$, a minimum of $0.706$, a maximum of $0.906$, as well as a standard deviation of $0.0365$. As mentioned, the percentile for validation was set to $0.9987$, while for test we used a percentile of $0.9981$. Post-processing parameters are optimized on validation, and accordingly set for test inference.

In Table~\ref{tab:results}, we highlight all individual models as well as the final blend with according settings, and validation and test scores. For all models, we employed the same thresholding, binary prediction adjustments, as well as post-processing, meaning that individual results only differ on the trained models used. Again, note that each individual model averages six different seeds to better account for individual randomness. For further details please also refer to respective configurations in our code base \footnote{See footnote \ref{githubrepo}}.

The results further highlight the difficulty of robust local validation generalizing well to the unseen test data. While we can observe a trend of higher CV leading to higher test scores, the best individual local validation models are not necessarily also the best models on public and private leaderboard scores. Also, while validation and public LB scores are more aligned, private LB scores exhibit significantly lower scores even exceeding the lower bounds of our bootstrapped validation scores as shown in Figure~\ref{fig:cv_results}. With test data being hidden, we can only speculate that some of the differences can be explained by the presence of two additional test sites (COL \& SNE) for which we can neither robustly judge the presence / absence of certain bird species nor the amount of bird segments in respective data strongly driving absolute scores. It is also unclear, whether public and private test data is split randomly, or based on a certain logic. However, note that we could gather these insights only after the competition ended, as we were only able to acquire limited insights into public LB scores (two submissions per day), and no insights about private LB during competition. Yet, these observations further strengthen the importance of robust solutions in this competition, that are able to generalize well to partly unseen data and locations. 

While we can observe some (partly random) fluctuations in individual model scores (although each one consists of six individual seeds), our final ensemble exhibits the best local validation score, best public LB score, and best private LB score. The ensemble successfully captures the weaknesses and strengths of each individual model and was a robust choice to generalize well on the unseen test data. A further testament of the robustness of our solution is given by the fact that we improved our rank on private leaderboard compared to our initial rank on public leaderboard compared to other contestants. The final leaderboard results are highlighted in Table~\ref{tab:lbresults}.



\section{Conclusions}

In this work, we discussed how to address challenges of using the crowd-sourced dataset xeno-canto for productive bird sound recognition
as part of the second place solution to BirdCLEF2021 \cite{birdclef2021}.  
Specifically, we showed how to efficiently finetune a pre-trained CNN with weakly labeled data by reshaping input and output tensors and how to handle noisy labels by using quality rating, as well as label smoothing. 
Furthermore, we demonstrated that mixing train dataset recordings internally as well as with external data significantly improves generalization to distinct acoustic environments. 
Thus, we gave suggestions to enhance autonomous monitoring of avian population which is essential for assessing the biodiversity and healthiness of ecosystems and a sustainable development of humanity.


\begin{acknowledgments}
We would like to thank Kaggle and the competition hosts for conducting this interesting competition as well as all other competitors on Kaggle for the challenge. Additionally, a big thank you goes to all users of xeno-canto generously uploading and sharing their bird recordings with the world. Without them this work would not have been possible.
\end{acknowledgments}

\footnotesize
\bibliography{literature}




\end{document}